\documentclass[nofootinbib,prapplied,superscriptaddress,longbibliography,a4paper,amsfonts,twocolumn]{revtex4-1}
\usepackage{graphicx}
\usepackage{natbib}
\usepackage{hyperref}
\usepackage{amsmath}

\newcommand{\ket}[1]{\ensuremath{\left| #1 \right\rangle}}

\newcommand{\ii}{\ensuremath{i}}

\usepackage[normalem]{ulem}
\usepackage[dvipsnames]{xcolor}

\begin{document}

\title{Fourier Transform of the Orbital Angular Momentum of a Single Photon}

\author{Jaroslav Kysela}
\affiliation{Institute for Quantum Optics and Quantum Information (IQOQI), Austrian Academy of Sciences, Boltzmanngasse 3, 1090 Vienna, Austria.}
\affiliation{Faculty of Physics, University of Vienna, Boltzmanngasse 5, 1090 Vienna, Austria.}

\author{Xiaoqin Gao}
\email{xiaoqin.gao@univie.ac.at}
\affiliation{National Mobile Communications Research Laboratory, Quantum Information Research Center, Southeast University, Sipailou 2, 210096 Nanjing, China.}
\affiliation{Institute for Quantum Optics and Quantum Information (IQOQI), Austrian Academy of Sciences, Boltzmanngasse 3, 1090 Vienna, Austria.}
\affiliation{Faculty of Physics, University of Vienna, Boltzmanngasse 5, 1090 Vienna, Austria.}

\author{Borivoje Daki{\'c}}
\email{borivoje.dakic@univie.ac.at}
\affiliation{Institute for Quantum Optics and Quantum Information (IQOQI), Austrian Academy of Sciences, Boltzmanngasse 3, 1090 Vienna, Austria.}
\affiliation{Faculty of Physics, University of Vienna, Boltzmanngasse 5, 1090 Vienna, Austria.}

\date{\today}

\begin{abstract}
Optical networks implementing single-qudit quantum gates may exhibit superior properties to those for qubits as each optical element in the network can work in parallel on many optical modes simultaneously. We present an important class of such networks that implements in a deterministic and efficient way the quantum Fourier transform (QFT) in an arbitrarily high dimension. These networks redistribute the initial quantum state into the orbital angular momentum (OAM) and path degrees of freedom and offer two modes of operation. Either the OAM-only QFT can be implemented, which uses the path as an internal auxiliary degree of freedom, or the path-only QFT is implemented, which uses the OAM as the auxiliary degree of freedom. The resources for both schemes scale linearly $O(d)$ with the dimension $d$ of the system, beating the best known bounds for the path-encoded QFT. While the QFT of the orbital angular momentum states of single photons has been applied in a multitude of experiments, these schemes require specially designed elements with non-trivial phase profiles. In contrast, we propose a different approach that utilizes only conventional optical elements. 
\end{abstract}

\maketitle

\section{Introduction}

The field of quantum computation has gained ever-increasing attention thanks to the invention of the quantum factoring algorithm due to Shor \cite{Shor1994, Ekert1996, Shor1997}, which utilizes as its key part a quantum Fourier transform. The quantum Fourier transform (QFT), or quantum Hadamard gate, has been since then used in many areas of quantum computation and communication \cite{Childs2010} for systems of qubits as well as high-dimensional qudits. The application areas of the high-dimensional QFT acting on a single photon's state include, but are not limited to, generation of mutually unbiased bases in the quantum state tomography \cite{Wootters1989, brierley2010, DURT2010, Giovannini2013} and quantum key distribution \cite{Groblacher2006, Malik2012}; generation of angular states \cite{Barnett1990, Franke-Arnold2004, Yao2006, Wang2017}; sorting of spatial modes of a photon \cite{Ionicioiu2016}; and representation of multiport devices employed in Bell test experiments \cite{Zukowski1997}. Single photon's high-dimensional QFTs can be also used as building blocks of programmable universal multi-port arrays \cite{lpezpastor2019arbitrary, Saygin2020, pereira2020universal}.

The orbital angular momentum (OAM) of single photons is a quantized property with infinite-dimensional Hilbert space, which allows for construction of qudits in arbitrarily high dimension \cite{Allen1992, Krenn2014, Erhard2018review}. By manipulation of a single photon's OAM the universal quantum computation is possible \cite{Garcia-Escartin2011, Gao2019}. The Fourier transform of the OAM eigenstates of single photons has been demonstrated in a number of experiments \cite{Lavery2012, OSullivan2012, Malik2012, Mirhosseini2013, Wang2017, Brandt2020} using free-space propagation and specially designed optical elements imparting non-trivial phase profiles. Alternative experimental schemes have been presented in special cases \cite{Song2015}.

In this paper we demonstrate a completely different general approach, which works with in principle 100~\% efficiency for arbitrarily high dimension of the OAM space of a single photon. Our scheme decomposes the Fourier transform into a series of elementary operations that can be directly implemented with basic commercially-available optical elements such as beam-splitters, mirrors and Dove prisms. Such an explicit decomposition reveals how individual components participate in the evolution of different OAM eigenstates and allows for modifications, such as miniaturization of the setup to a micro-chip level. The scheme's implementation is recursive and makes use of an interplay between the OAM and path degrees of freedom. The number of required optical elements $O(d)$ scales linearly in the dimension $d$, as opposed to $O(d \log d)$ scaling of the setup using path-encoded qudits \cite{Reck1994, Torma1996, Barak2007}. This is made possible by the fact that a single passive optical element can act on many OAM eigenstates at the same time, leading to a heavily parallelized operation of the network of optical elements. Moreover, the setup for the OAM Fourier transform can be modified to act as the path-only Fourier transform. In such a scheme, the OAM is present only in the inner workings of the transform. This OAM-enhanced setup preserves the linear scaling of the number of optical elements, which shows a clear advantage of our scheme over the setup that uses only the path degree of freedom.

One of our scheme's main components is the OAM-Path swap operator, which interchanges the OAM and path degrees of freedom of a photon's state. To the best of our knowledge, we demonstrate for the first time the implementation of such an operator in terms of conventional optical elements. The OAM-Path swap represents a multiport generalization of the OAM sorter and its implementation features efficient deployment of the OAM parity sorter. Each instance of the parity sorter functions simultaneously as a series of many conventional beam-splitters for different OAM eigenstates.

The manuscript is organized as follows. In section \ref{sec:theory} we introduce the theoretical background for the construction of the Fourier transform. Then we present in section \ref{sec:implementation} the setup that implements the Fourier transform in the OAM of a single photon. We discuss the properties of the OAM sorter, a key part of the setup, in section \ref{sec:sorter}. In section \ref{sec:swap}, we demonstrate how to generalize the OAM sorter into the OAM-Path swap operator. In section \ref{sec:scaling} the scaling of our scheme is presented. In section \ref{sec:path}, we compare our scheme with the recursive scheme for the path-only Fourier transform and summarize our results in the last section \ref{sec:conclusion}.

\begin{figure}
    \centering
    \includegraphics[width=0.47\textwidth]{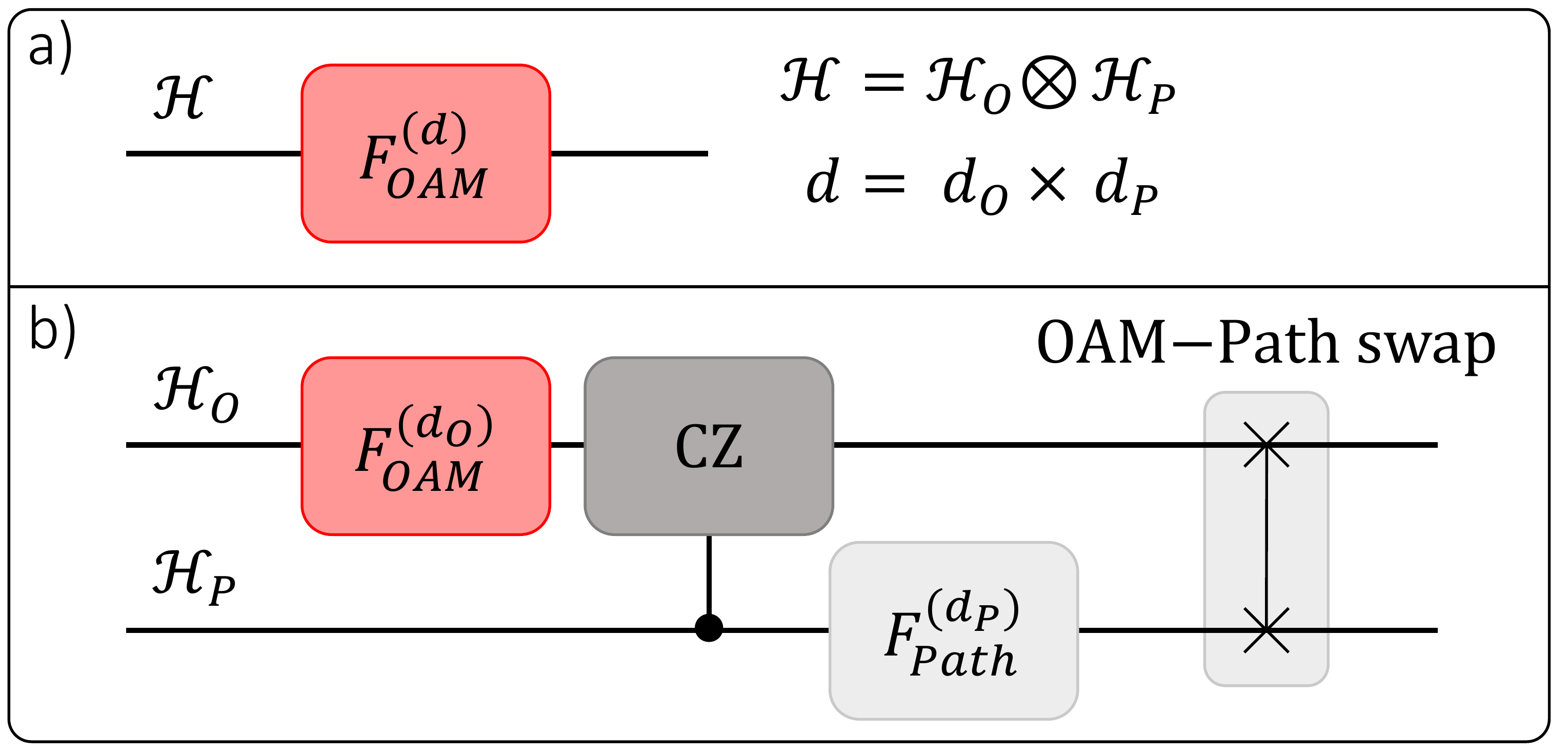}
    \caption{Circuit representation of the recursive scheme for the OAM Fourier transform. a) The OAM Fourier transform acting on the $d$-dimensional OAM space $\mathcal{H}$, which can be decomposed into a tensor product of two factor subspaces $\mathcal{H}_O$ and $\mathcal{H}_P$. b) A circuit equivalent to a), where a $d_O$-dimensional OAM Fourier transform is applied first, followed by a phase gate $\mathrm{CZ}$ and a $d_P$-dimensional path-only Fourier transform. The swap operator then exchanges states between the two subspaces. For details refer to the main text.}
    \label{fig:circuit_fft}
\end{figure}

\begin{figure*}
    \centering
    \includegraphics[scale=.45]{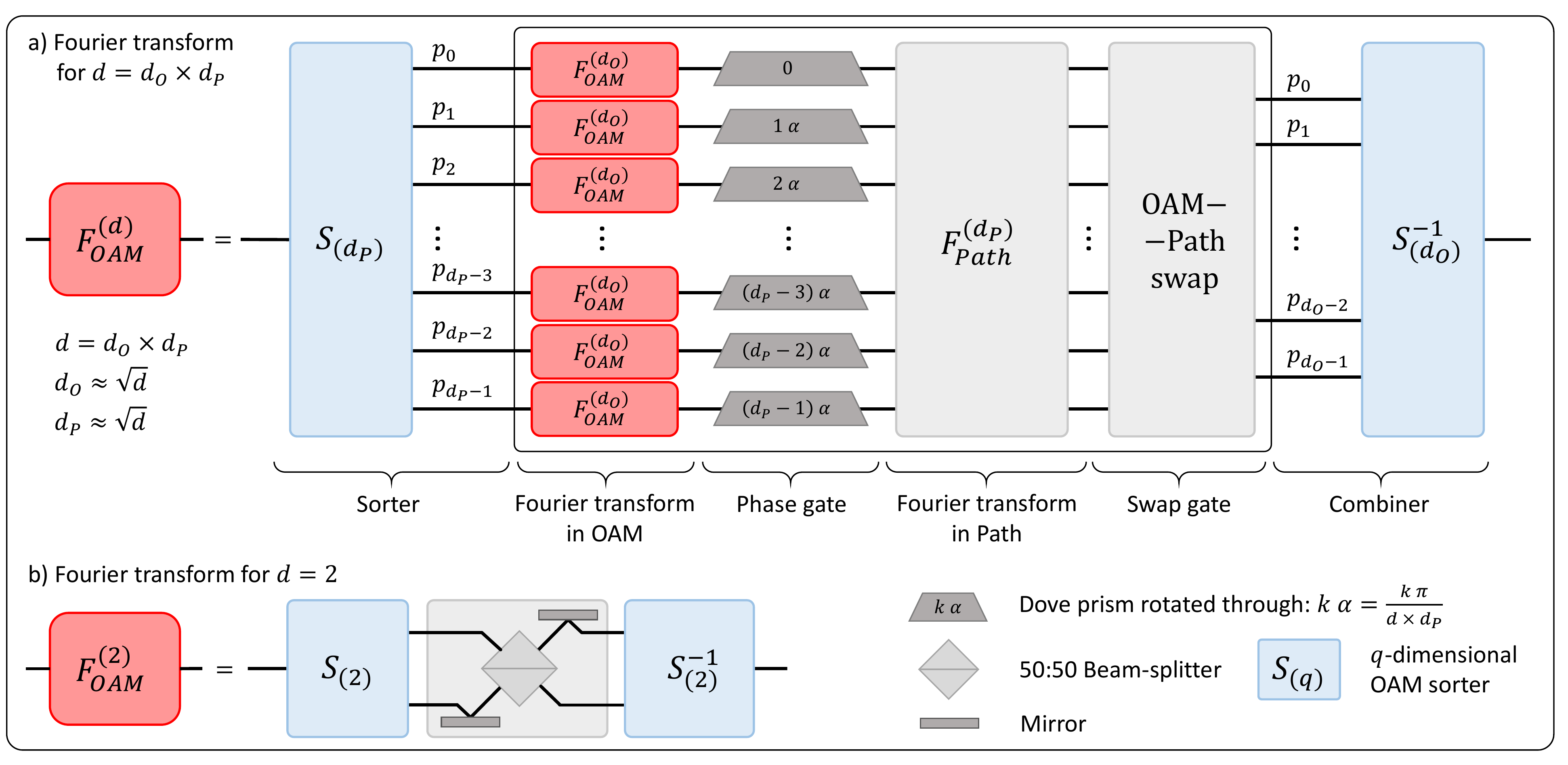}
    \caption{General scheme of the Fourier transform in the orbital angular momentum (OAM) of single photons. a) The Fourier transform in dimension $d = d_O \times d_P > 2$ is constructed recursively making use of Fourier transforms in smaller dimensions $d_O$ and $d_P$. The OAM eigenstates in the initial quantum superposition are at first redistributed by an OAM sorter $S_{(d_P)}$ such that along each of the $d_P$ different paths denoted by $p_0$ through $p_{d_P-1}$ there propagate OAM eigenstates of the form $\ket{0}$, $\ket{d_P}$, \ldots, $\ket{d_P(d_O-1)}$. The Fourier transform itself is then performed in four steps. The first step consists in the application of the $d_O$-dimensional Fourier transform $F_{OAM}^{(d_O)}$ on each path. Then a phase gate is applied on all paths, which is implemented by a series of Dove prisms. In the third step a single $d_P$-dimensional path-only Fourier transform $F_{Path}^{(d_P)}$ is applied on the path degree of freedom for all OAM eigenstates. All beam-splitters in the path-only Fourier transform are supplemented by two mirrors as demonstrated in b). The fourth step is represented by the OAM-Path swap gate. Finally, all states are recombined into a single output path by the second OAM sorter $S^{-1}_{(d_O)}$, which is operated in reverse.  b) The elementary building block of the recursive scheme -- Fourier transform for $d=2$ -- consists of two OAM sorters and a single beam-splitter, which is complemented by two mirrors such that the OAM value of the incoming eigenstates is not affected by the reflection off the beam-splitter's interface.}
    \label{fig:fft}
\end{figure*}

\section{Fourier transform}
\label{sec:theory}

In this section, we present a recursive scheme for the construction of the Fourier transform that acts on the OAM of a single photon. The initial state of a photon is a superposition of OAM eigenstates $\sum_{j=0}^{d-1} \alpha_j \ket{j}$. The Fourier image of such a state is then a superposition $\sum_{k=0}^{d-1} \beta_k \ket{k}$, where coefficients $\beta_k$ satisfy
\begin{equation}
    \beta_k \equiv \frac{1}{\sqrt{d}}\sum_{j=0}^{d-1} e^{\ii \frac{2 \pi}{d} j k} \alpha_j.
    \label{eq:coef_formula_fft}
\end{equation}
Our implementation of the Fourier transform is inspired by the classical fast Fourier transform algorithms \cite{Cooley1965}, an idea already applied in the field of quantum information for the path degree of freedom \cite{Torma1996, Zhang2006, Barak2007, Tabia2015}. 

To implement the $d$-dimensional Fourier transform in the OAM, we first decompose the total $d$-dimensional OAM space of a photon into a tensor product $\mathcal{H} = \mathcal{H}_{O} \otimes \mathcal{H}_P$, as shown in Fig.~\ref{fig:circuit_fft} a). There, a $d_O$-dimensional subspace $\mathcal{H}_O$ is spanned by a subset of original OAM eigenstates and $\mathcal{H}_P$ is a $d_P$-dimensional subspace represented by the path degree of freedom, such that $d = d_O \times d_P$. Henceforth we denote a quantum state of a photon with $m$ quanta of OAM by $\ket{m}_O$ and call it an OAM \emph{eigenstate}. OAM eigenstates represent a basis of the subspace $\mathcal{H}_{O}$. Likewise, a basis vector $\ket{l}_P$ of the subspace $\mathcal{H}_{P}$ is called a \emph{propagation mode} or simply a \emph{path}. For each division of dimension $d$ into a pair of smaller dimensions $d_O$ and $d_P$ there is a one-to-one correspondence between an index $j \in \{0, \ldots, d-1\}$ and a pair of numbers $(m, l)$ with $0 \leq m < d_O$ and $0 \leq l < d_P$ such that
\begin{eqnarray}
\ket{m}_O\ket{l}_P & \quad \Leftrightarrow \quad & \ket{j} = \ket{m \times d_P + l}. \label{eq:tensor_index}
\end{eqnarray}
In cases when $d$ is a prime number the decomposition into $d_O$ and $d_P$ is trivial and the recursive scheme presented below is not applicable. For these dimensions a different approach has to be chosen and we discuss one possible alternative in the following section.

The Fourier transform in dimension $d$ is obtained in four steps as demonstrated in Fig.~\ref{fig:circuit_fft} b). In the first step, a $d_O$-dimensional Fourier transform is applied only on the OAM subspace $\mathcal{H}_O$. Then a controlled phase gate acts on both subspaces, followed by a $d_P$-dimensional Fourier transform applied only on the path subspace. As the last step, a swap gate is used, which effectively exchanges the OAM and path subspaces. This procedure can be mathematically summarized by the formula
\begin{equation}
    F^{(d)}_{OAM} = \textrm{SWAP} \cdot F^{(d_P)}_{Path} \cdot \textrm{CZ} \cdot F^{(d_O)}_{OAM},
    \label{eq:formula_fft}
\end{equation}
where the phase gate $\textrm{CZ}$ acts as a high-dimensional controlled-Z gate on the OAM degree of freedom
\begin{equation}
    \textrm{CZ}(\ket{m}_O\ket{l}_P) = (Z^{l}\ket{m}_O)\ket{l}_P = \omega^{m \times l} \ket{m}_O \ket{l}_P
    \label{eq:formula_phase_gate}
\end{equation}
with $\omega = \exp{(2 \pi \ii/d)}$. The action of the swap on an input mode reads $\textrm{SWAP}(\ket{r}_O\ket{q}_P) = \ket{q}_O\ket{r}_P$. Note that when $d_O \neq d_P$ the swap operator effectively changes dimensions of the two subspaces $\mathcal{H}_O$ and $\mathcal{H}_P$. This imposes nevertheless no restrictions on our implementation. The resulting mode $\ket{q}_O\ket{r}_P$ can be now identified with a single index $\ket{k}$ in a way analogous to Eq.~\ref{eq:tensor_index} as
\begin{eqnarray}
\ket{q}_O\ket{r}_P & \quad \Leftrightarrow \quad & \ket{k} = \ket{q \times d_O + r}.
\label{eq:tensor_index_rev}
\end{eqnarray}
At the end, one obtains the Fourier image of the OAM eigenstates of a single photon leaving the device along a single output path. For details see Appendix \ref{sec:decomposition}.

We have thus demonstrated how to construct a high-dimensional Fourier transform using lower-dimensional Fourier transforms that act on the OAM and path degrees of freedom. In the following section we present the corresponding implementation scheme.

\section{Implementation}
\label{sec:implementation}

The general recursive scheme based on the decomposition in Eq.~\eqref{eq:formula_fft} works in principle for any dimension $d$, provided that $d$ is not a prime number. Nevertheless, from now on we restrict ourselves only to dimensions that are powers of two, i.e. $d = 2^M$ for some $M \in \mathbb{N}$. For these dimensions an efficient implementation can be found for each component of the general scheme. We discuss the case with $d \neq 2^M$ at the end of this section.

The efficient recursive scheme that implements the Fourier transform is depicted in Fig.~\ref{fig:fft} a) and comprises six separate modules. Each module is described in detail below.

\begin{enumerate}
    \item In the first module, the initial superposition of $d$ OAM eigenstates is split such that there is a smaller number $d_O$ of OAM eigenstates propagating along each of $d_P$ different paths. This splitting, corresponding to relabeling $\ket{j} \to \ket{m}_O\ket{l}_P$ in Eq.~\ref{eq:tensor_index}, is performed by the $d_P$-dimensional OAM sorter. The structure and operation of the OAM sorter are described in detail in the following section.

    \item The second module consists of a collection of identical $d_O$-dimensional OAM Fourier transforms, each of which is applied onto a different path to implement $F^{(d_O)}_{OAM}$ in Eq.~\ref{eq:formula_fft}. The $d_O$-dimensional Fourier transform itself is constructed recursively following the same pattern as the one presented in this section when one replaces $d$ with the value of $d_O$. The elementary building block -- the Fourier transform for $d=2$ -- is depicted in Fig.~\ref{fig:fft} b).

    \item The third module, which represents the phase gate $\textrm{CZ}$ in Eq.~\ref{eq:formula_phase_gate}, is implemented by a series of properly rotated Dove prisms. They impart additional phases to OAM eigenstates that pass through them.

    \item The fourth module comprises a single path-only $d_P$-dimensional Fourier transform, which performs transformation $F^{(d_P)}_{Path}$ in Eq.~\ref{eq:formula_fft}. This transform affects only the path degree of freedom and acts identically onto each OAM eigenstate. The implementation of the path-only Fourier transform in terms of beam-splitters and phase-shifters is given in Refs.~\cite{Torma1996, Barak2007}. To compensate for the OAM inversion $\ket{m}_O \to \ket{-m}_O$ due to reflection off the beam-splitter's interface, each beam-splitter in the path-only Fourier transform has to be complemented by two extra mirrors as demonstrated in Fig.~\ref{fig:fft} b).

    \item The fifth module is the $\textrm{SWAP}$ gate which reorders the coefficients of the joint OAM-Path state such that the coefficient for mode $\ket{m}_O\ket{l}_P$ becomes the coefficient for mode $\ket{l}_O\ket{m}_P$. The swap gate, whose structure and working principle are one of the main results of this paper, is described in detail in a separate section. Without the swap gate, the OAM sorter in the sixth module would also perform undesirable additional permutation of the output OAM eigenstates.

    \item In the sixth module, all OAM eigenstates are rerouted into a single output path by the $d_O$-dimensional OAM sorter, which is operated in reverse. This action corresponds to the relabeling in Eq.~\ref{eq:tensor_index_rev}.
\end{enumerate}

It is important to note that OAM eigenstates entering the lower-dimensional OAM Fourier transforms are of the form $\ket{0}_O$, $\ket{d_P}_O$, \ldots, $\ket{d_P(d_O-1)}_O$ (in the first recursion), so the difference between two successive OAM eigenstates, or multiplicity, is $d_P$. This fact has to be reflected in the order of OAM exchangers (see the next section) and rotation of Dove prisms in the lower-dimensional Fourier transforms. Specifically, the order of all exchangers has to be multiplied by $d_P$ and the angle of rotation for all Dove prisms has to be divided by $d_P$ (compare also the form of angle $\alpha$ in Fig.~\ref{fig:fft}). With each recursion the multiplicity of input eigenstates increases correspondingly.

The recursive scheme in Fig.~\ref{fig:fft} requires a number of optical elements that scales linearly with the dimension, as is shown in section \ref{sec:scaling}. This scaling is made possible by the efficient implementation of OAM sorters, swap operators and the path-only Fourier transforms. Such an implementation is nevertheless available only for dimensions of the form $d = 2^M$. For dimensions that are not a power of two alternative implementations exist, but these may require asymptotically more resources. The extreme case is when $d$ is a prime number, where the known recursive schemes cannot be used. The brute-force alternative to our recursive scheme, which works for any dimension, consists of three steps. In the first step, a $d$-dimensional OAM sorter transforms the initial OAM eigenstates into propagation modes. In the second step, a $d$-dimensional path-only Fourier transform is applied, which can be implemented using Reck et al. design \cite{Reck1994}. In the third step, the resulting propagation modes are transformed back to the OAM with the help of another OAM sorter, which is operated in reverse. The number of elements in such a brute-force approach scales quadratically with the dimension. It is still an open question how to implement efficiently the scheme in Fig.~\ref{fig:circuit_fft} for a general dimension $d \neq 2^M$.

\begin{figure}[t]
    \centering
    \includegraphics[scale=.4]{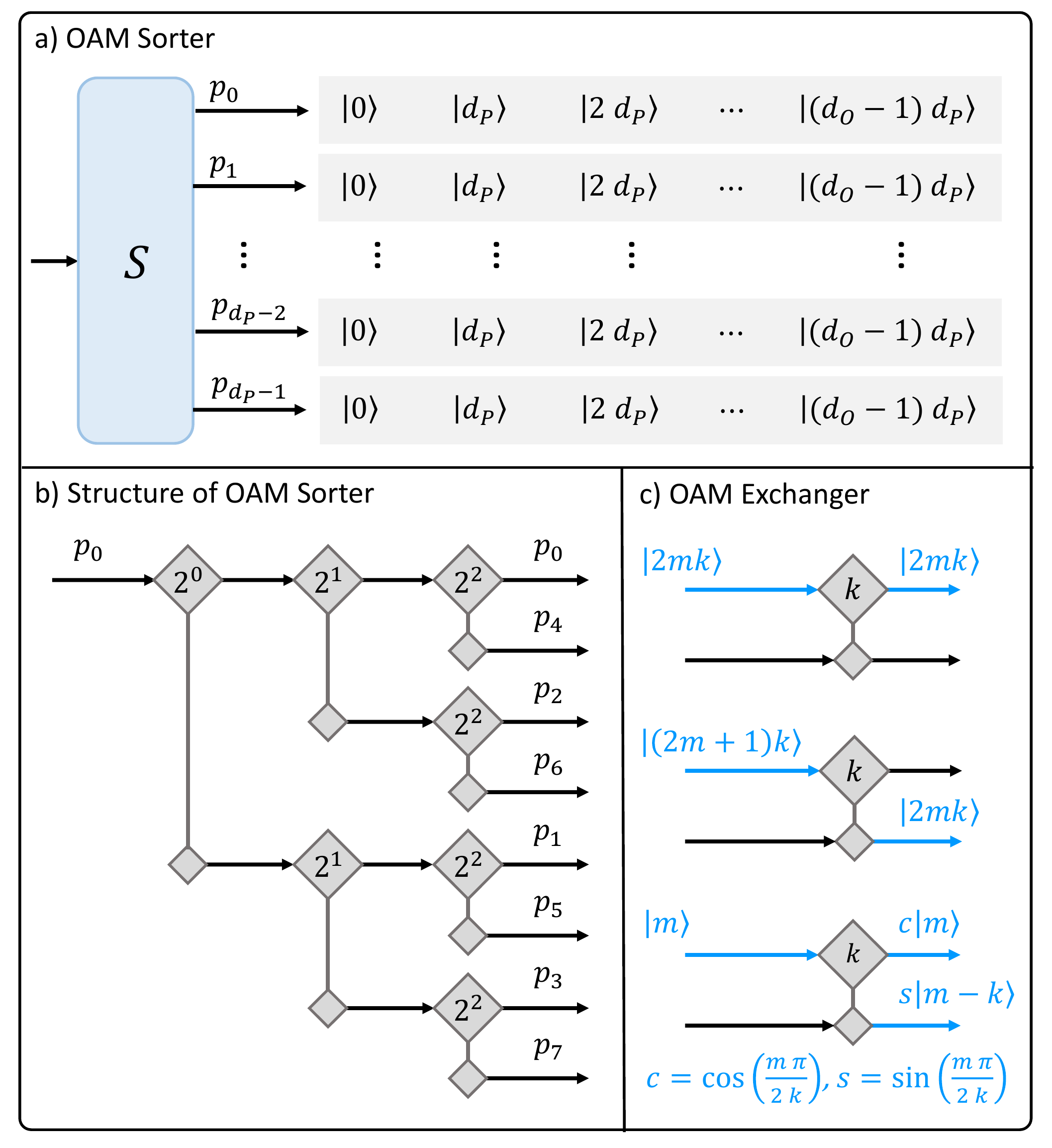}
    \caption{OAM Sorter. a) The OAM sorter for a general power-of-two dimension $d_P$. All OAM eigenstates from the subspace $\{\ket{0}_O, \ldots, \ket{d_P-1}_O\}$ get transformed into the eigenstate $\ket{0}_O$ propagating in different paths. Eigenstates from the subspace $\{\ket{d_P}_O, \ldots, \ket{2 \, d_P-1}_O\}$ are sorted analogously, but the resulting OAM eigenstate is $\ket{d_P}_O$. This modulo property holds for arbitrarily large OAM values of input eigenstates. b) The binary-tree-like structure of OAM exchangers with an increasing order works as a sorter of OAM eigenstates. Here, a specific example for $d_P = 8$ is shown. c) The OAM exchanger $\textrm{EX}_k$ is a composition of two holograms and the elementary OAM parity sorter \cite{Leach2002}. OAM eigenstates that are even multiples of the order $k$ of the exchanger and enter its upper port leave the output upper port unaffected. The odd multiples entering the upper port are rerouted to the lower output port and loose $k$ quanta of OAM. Other OAM eigenstates that are not multiples of $k$ leave the exchanger in a superposition of eigenstates and paths. A single OAM exchanger can thus work as an identity, a switch and a beam-splitter for different input eigenstates. For eigenstates entering the lower input port the exchanger works analogously.}
    \label{fig:OAM_sorter_exchanger}
\end{figure}

\section{OAM Sorter}
\label{sec:sorter}

The first stage of the setup for the Fourier transform consists of an OAM sorter. The OAM sorter is a device that transforms the OAM eigenstates of incoming photons into different propagation modes. For a fixed dimension $d_P$ the sorter sorts input eigenstates $\ket{0}_O$ through $\ket{d_P-1}_O$, which enter the first path $p_0$, into separate output paths $p_0$ through $p_{d_P-1}$. All such input eigenstates leave the sorter in OAM eigenstate $\ket{0}_O$. Various designs of the sorter have been realized using e.g. the multi-plane light conversion \cite{Fontaine2019,Labroille2014}, light scattering in random media \cite{Fickler2017} or light propagation through two specially designed phase plates \cite{Berkhout2010}. Throughout this paper we consider the design due to Leach et al. \cite{Leach2002, Leach2004, Garcia-Escartin2008, Erhard2017, Xgate} and discuss alternative designs in section \ref{sec:scaling}. This design has the following modulo property. When an OAM eigenstate $\ket{m}_O$ with $m \ge d_P$ is injected into the sorter, its propagation is analogous to that of OAM eigenstates with $m < d_P$ except that the output eigenstate is no longer $\ket{0}_O$. It turns out that the input OAM eigenstate $\ket{m}_O \ket{0}_P$ gets transformed according to relations
\begin{equation}
    S_{(d_P)}(\ket{m}_O\ket{0}_P) = \ket{\, d_P \left\lfloor\frac{m}{d_P} \right\rfloor}_O\ket{\ \rule{0ex}{3ex}m \ \mathrm{mod} \ d_P \,}_P,
\end{equation}
where $\lfloor x \rfloor$ is the integral part of $x \in \mathbb{R}$ and $S_{(d_P)}$ denotes a $d_P$-dimensional OAM sorter. This modulo property is illustrated in Fig.~\ref{fig:OAM_sorter_exchanger}~a) and corresponds exactly to the relabeling introduced in Eq.~\ref{eq:tensor_index}. 

The sorter is constructed as a binary-tree network of OAM-manipulating elements in a way shown in Fig.~\ref{fig:OAM_sorter_exchanger} b). Each of these elements, henceforth referred to as \emph{OAM exchangers}, is an interferometric device composed of an OAM parity sorter \cite{Leach2002, Erhard2017} and two holograms, which shift the OAM value of the input eigenstate \cite{Garcia-Escartin2008}. Note that additional permutation of paths is necessary in Fig.~\ref{fig:OAM_sorter_exchanger} b) to comply with the order of paths depicted in Fig.~\ref{fig:OAM_sorter_exchanger} a). The OAM exchanger $\textrm{EX}_k$ exhibits three modes of operation based on its order $k$ and the state of the incoming photon, see Fig.~\ref{fig:OAM_sorter_exchanger} c). A photon in OAM eigenstate $\ket{m \, k}$ that enters the upper port of the exchanger $\textrm{EX}_k$ leaves its upper or lower output port depending on the parity of $m \in \mathbb{Z}$. All other input OAM eigenstates, which are not multiples of $k$, leave the exchanger in a superposition of both output ports. A single exchanger therefore behaves either as an identity, or as a switch, or as a beam-splitter with varying splitting ratio for different OAM eigenstates. This beam-splitter-like property was first utilized in Ref.~\cite{Gao2019CNOT} for the special case when the order of the exchanger is $k = 2$. In this paper we show that a single exchanger of order $k$ works effectively as $4 k$ different beam-splitters simultaneously. For a more detailed description of the exchanger refer to Appendix \ref{sec:optical_elements}. 

\begin{figure}[htb]
    \centering
    \includegraphics[scale=.45]{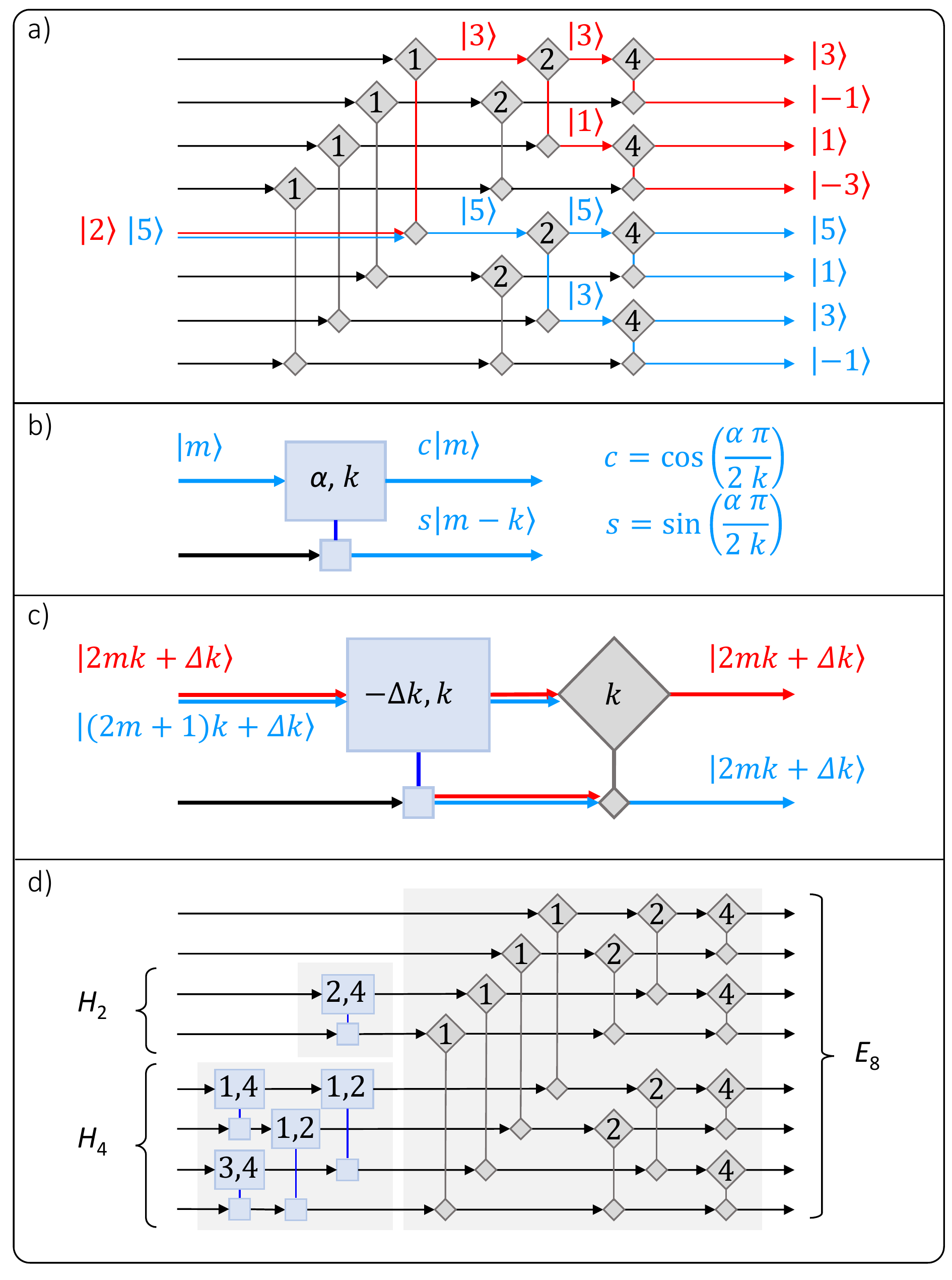}
    \caption{OAM-Path swap operator. a) The network of OAM exchangers with increasing orders of the form $2^j$ represents a naive generalization of the OAM sorter. By propagation through this network, superpositions of OAM eigenstates and paths are introduced into the output state as exemplified for eigenstates $\ket{2}$ and $\ket{5}$ entering the fifth input port. b) Such an undesirable behavior can be counteracted by adding a collection of holo-beam-splitters into the network. The holo-beam-splitter $\textrm{HBS}_{(\alpha, k)}$ of order $(\alpha, k)$ is a device consisting of a conventional beam-splitter with a splitting ratio $\alpha \pi/(2 k)$ and two holograms of opposite values $k$ and $-k$. For further details refer to Appendix \ref{sec:optical_elements}. c) The crucial feature of the holo-beam-splitter is that it effectively shifts the splitting properties of the OAM exchangers, such that eigenstates $\ket{m k + \Delta k}$, that would otherwise leave the exchanger of order $k$ in a superposition, leave in only one of the two output ports of the exchanger. d) As demonstrated in Appendix \ref{sec:structure_swap}, the resulting network implementing the OAM-Path swap comprises the network of exchangers $E_{d_P}$ together with the series of gradually larger networks $H_j$ of holo-beam-splitters. In the figure a special case of a swap operator for $d_P = d_O = 8$ is shown.}
    \label{fig:leach_plus_bs}
\end{figure}

\section{OAM-Path Swap}
\label{sec:swap}

In the final part of the setup of the Fourier transform, the swap operator is utilized. Its operation on individual modes can be summarized as
\begin{equation}
    \textrm{SWAP}(\ket{m}_O\ket{l}_P) = \ket{l}_O\ket{m}_P,
    \label{eq:oam_transposition}
\end{equation}
where $\ket{m}_O\ket{l}_P$ denotes an OAM eigenstate $0 \le m < d_O$ in path $p_l$. Analogously to the OAM sorter discussed in the previous section, also the OAM-Path swap operator exhibits the modulo property for $m \ge d_O$.

The swap operator can be understood as a generalization of the OAM sorter \cite{Ghiu2019}. One could naively expect that a complete network of OAM exchangers of increasing orders $2^0$ to $2^{d_P-1}$ works as a swap, see the example for $d_P = 8$ in Fig.~\ref{fig:leach_plus_bs} a). Unfortunately, such a network not only permutes the input eigenstates as it is supposed to, but also gives rise to superpositions of multiple eigenstates and paths. When an OAM eigenstate is injected to any of the input ports $p_l$ of such a network, where $l \ge 3$, it leaves the network in a superposition of as many as $d_P/2$ different modes. This is caused by the fact that at some point of its evolution through the network the initial eigenstate attains the form $\ket{m \, k + \Delta k}_O$ with $0 < \Delta k < k$. The OAM value of this eigenstate is not a multiple of the order of the OAM exchangers. From that point onward, all exchangers work on such an eigenstate as beam-splitters with varying splitting ratios and the output state is thus a superposition of different eigenstates in different paths. Specifically, for eigenstate $\ket{2 \, m \, k + \Delta k}_O$ entering the upper port $p_{a}$ of the OAM exchanger $\textrm{EX}_k$ of order $k$ one obtains (omitting a global phase)
\[\textrm{EX}_k(\ket{2 m k + \Delta k}_O \ket{a}_P) = \hspace{4cm} \] \\[-7ex]
\begin{eqnarray}
& & \cos{\left(\frac{\pi \Delta k}{2 k}\right)} \ket{2 m k + \Delta k}_O \ket{a}_P \nonumber \\
& + & \sin{\left(\frac{\pi \Delta k}{2 k}\right)} \ket{(2 m - 1) k + \Delta k}_O \ket{b}_P,
\label{eq:exch}
\end{eqnarray}
and analogously for eigenstates $\ket{(2 m + 1) \, k + \Delta k}_O$ and the lower input port $p_{b}$. The exchanger $\textrm{EX}_k$ in this specific case therefore works effectively as a beam-splitter with the splitting ratio equal to $\pi \Delta k/(2 k)$.

The undesirable emergence of superpositions is avoided when we augment the network with a collection of \emph{holo-beam-splitters}. A holo-beam-splitter $\textrm{HBS}_{(\alpha, k)}$ of order $(\alpha, k)$ is a passive optical device consisting of a conventional beam-splitter with a splitting ratio $\alpha \pi/(2 k)$ and two holograms of opposite values $k$ and $-k$. Its operation on eigenstates entering its upper port is summarized in Fig.~\ref{fig:leach_plus_bs} b) and its detailed structure is presented in Appendix \ref{sec:optical_elements}. A crucial observation is that one can force the OAM exchanger of order $k$ to work as a mere identity or switch even for eigenstates $\ket{m \, k + \Delta k}_O$ that are not multiples of $k$. One can do so by prepending a holo-beam-splitter of order $(-\Delta k, k)$ to the exchanger, as demonstrated in Fig.~\ref{fig:leach_plus_bs} c) \cite{[{This trick is somewhat similar to the use of an extra phase plate in the OAM parity sorter as reported in: }]Wei2003}. We obtain transformation rules (again omitting a global phase)
\begin{multline}
    (\textrm{EX}_k \cdot \textrm{HBS}_{(-\Delta k, k)}) (\ket{2 m k + \Delta k}_O \ket{a}_P) = \\ \ket{2 m k + \Delta k}_O \ket{a}_P,
\end{multline}
where analogous relations hold also for eigenstates $\ket{(2 m + 1) \, k + \Delta k}_O$ and the lower input port $p_{b}$. In this formula the holo-beam-splitter and the exchanger act on the same states as in Eq.~\ref{eq:exch}, but this time no superposition emerges.

One can stack multiple setups in Fig.~\ref{fig:leach_plus_bs} c) to create a larger network. This way we arrive at the setup that implements an OAM-Path swap operator for general power-of-two dimensions $d_O$ and $d_P$. The swap gate consists of a network of exchangers and a series of networks of holo-beam-splitters of increasing size as shown in Fig.~\ref{fig:leach_plus_bs} d) for the case of $d_O = d_P = 8$. For the detailed explanation of the construction and structure of the swap operator refer to Appendix \ref{sec:structure_swap}. The resulting network works as a proper sorter for \emph{all} input ports, where the OAM value of output eigenstates contains information about the path into which the original eigenstate was injected.

\section{Scaling properties}
\label{sec:scaling}

In real-world applications it is necessary to assess the effect of imperfections and noise on the stability and overall feasibility of the setup. The stability of the scheme is the more of concern the more interferometers are employed. Their number depends on the number of beam-splitters. In our scheme the original $d$-dimensional Hilbert space is decomposed into a product of two subspaces of dimensions $d_O$ and $d_P$, such that $d=d_O \times d_P$. The values of $d_O$ and $d_P$ are nevertheless not fixed. One can search for such a combination of $d_P$ and $d_O$ that minimizes the number of beam-splitters in the setup of the Fourier transform, while still satisfying $d=d_O \times d_P$. Our simulations show that the optimal number of beam-splitters scales approximately linearly as $6.037 \times d$. This optimal scenario tends to prefer choices with $d_P \approx d_O$. In Appendix \ref{sec:scaling_high_dimensions} the linear scaling is analytically confirmed for a subset of dimensions of the form $d = 2^{2^M}$ with $M \in \mathbb{N}$, for which $d_P = d_O$ is used. Note that the logarithmic scaling of the Fourier transform setup reported in Ref.~\cite{Gao2019} relates to the number of elementary gates, not actual optical elements. When implementing the proposal in Ref.~\cite{Gao2019} with beam-splitters, their number scales as $O(d (\log_2(d))^5)$. 

Apart from the stability, the performance of the setup is also negatively affected by the losses introduced by beam-splitters and other optical elements. It turns out that the number of these additional optical elements also scales linearly in the dimension $d$. For the optimal scenario described above, one needs approximately $4.860 \times d$ Dove prisms, $3.037 \times d$ holograms and $2.976 \times d$ phase-shifters. For details see Appendix \ref{sec:scaling_high_dimensions}. The linear scaling of the whole scheme is made possible by an efficient implementation of the swap gate, which requires approximately $3 d \log_2(d)/2$ beam-splitters. There are alternative brute-force implementations of the swap gate, but these require asymptotically larger number of beam-splitters.

The OAM sorter design we adopted is the one due to Leach et al. \cite{Leach2002}, which is based on a network of interferometers. Our scheme makes use of the modulo property of this specific design in each recursion. Nevertheless, after all the recursions are performed, the initial part of the final setup represents a single $d$-dimensional sorter, where the modulo property plays no role any more. One can therefore replace this initial part with a sorter built using alternative designs. For example, one could use the design of Ref.~\cite{Berkhout2010} that uses only two plates with special phase-profiles. This way one can save around $2 d$ beam-splitters and equal number of other elements, such that the scheme then requires approximately $4 d$ beam-splitters.

\section{OAM vs. Path}
\label{sec:path}

The proposed scheme makes use of an interplay between the OAM and path degrees of freedom to efficiently perform the Fourier transform in the OAM degree of freedom. No other properties of incoming photons, such as the polarization, are affected. Nevertheless, the scheme can be after slight modification used also as a path-only Fourier transform, where the OAM degree of freedom plays the role of an intermediary that does not appear either in the input or the output state. Specifically, the first part of the recursive scheme, see Fig.~\ref{fig:fft} a), represented by a series of OAM sorters of decreasing dimensions, can be removed completely. We are then left with $d$ input ports. The last part of the scheme has to be adjusted by removing the very last reverted sorter and adding a series of additional sorters to obtain $d$ output ports. This OAM-enhanced scheme of the path-only Fourier transform shows better scaling properties in terms of the number of beam-splitters than the scheme presented in Refs.~\cite{Torma1996, Barak2007}. The OAM-enhanced scheme requires $O(d)$ beam-splitters as opposed to $O(d \log_2(d))$ beam-splitters of the original scheme with the crossing point from which OAM-enhanced scheme prevails occurring at $d=8192=2^{13}$. This improvement is made possible by optimal redistribution of coefficients of the quantum state between the OAM and path degrees of freedom.

\section{Conclusion}
\label{sec:conclusion}

We proposed a scheme for an efficient implementation of the Fourier transform that acts on the orbital angular momentum of single photons. Only commercially accessible optical elements are used in our scheme. An integral component of the scheme is the OAM-Path swap operator, which is a generalization of the OAM sorter for multiple input ports. In its implementation a heavy use is made of non-trivial parallel operation of OAM exchangers, which are elementary building blocks of the OAM sorter. A single exchanger, a passive element composed among others of two conventional beam-splitters, can work as many beam-splitters with varying splitting ratio at the same time. This property may be used in a more general framework, where each OAM eigenstate undergoes a different complex, yet precisely tailored, evolution by propagating through the identical network of standard optical elements. Even though algorithms for the decomposition of a general unitary into separate gates for path and OAM degrees of freedom exist \cite{Dhand2015, DeGuise2018, kumar2020optimal}, our scheme is to our knowledge the first explicit example of such a network.

The number of optical elements required in our scheme scales as $O(d)$, which is an improvement over the scaling $O(d \log_2(d))$ of the Fourier transform setup acting on path-encoded qudits \cite{Torma1996, Barak2007}. Our scheme can be after a slight modification used also to implement the path-only Fourier transform while preserving the scaling properties $O(d)$. The modulo property of the scheme allows one to use different $d$-dimensional OAM subspaces, such as $\{\ket{-d/2+1}, \ldots, \ket{d/2}\}$, which is naturally produced in the process of parametric down-conversion and which imposes less stringent requirements on the precision of the OAM-manipulating elements.

\begin{acknowledgments}
The authors thank Anton Zeilinger, Mario Krenn, Manuel Erhard, Armin Hochrainer, Marcus Huber, Robert Fickler and Elizabeth Agudelo for valuable discussions. XQG thanks Bin Sheng and Zaichen Zhang for support. This work was supported by the Austrian Academy of Sciences (OeAW), the European Research Council (SIQS Grant No. 600645 EU-FP7-ICT), and the Austrian Science Fund (FWF): F40 (SFB FoQuS) and W 1210-N25 (CoQuS). XQG acknowledges support from the National Natural Science Foundation of China (No. 61501109). BD acknowledges support from an ESQ Discovery Grant of the Austrian Academy of Sciences (OAW) and the Austrian Science Fund (FWF) through BeyondC (F71).

JK and XQG contributed equally to this work.
\end{acknowledgments}

\appendix

\section{Decomposition of the Fourier Transform}
\label{sec:decomposition}

In this section we prove the validity of the decomposition formula in Eq.~\ref{eq:formula_fft} by calculating explicitly the action of this decomposition on input OAM eigenstates. At first we relabel the input OAM eigenstate $\ket{j}$ in accord with the main text (cf. Eq.~\ref{eq:tensor_index}) as $\ket{j} = \ket{m \, d_P + l} = \ket{m}_O\ket{l}_P$. Then the application of the OAM Fourier transform on the first ket yields
\begin{equation}
    \ket{j} = \ket{m}_O\ket{l}_P \to \frac{1}{\sqrt{d_O}}\sum_{r=0}^{d_O-1} e^{\ii \frac{2 \pi}{d_O} m r} \ket{r}_O\ket{l}_P.
    \label{eq:fft_first_step}
\end{equation}
The phase gate multiplies each mode $\ket{r}_O\ket{l}_P$ by $\exp{(2 \pi \ii r l/d)}$ and the swap gate exchanges the two kets. After the sequential application of the phase gate, the path-only Fourier transform, and the swap gate the state on the right-hand side of Eq.~\ref{eq:fft_first_step} transforms into
\begin{equation}
    \frac{1}{\sqrt{d_O d_P}}\sum_{r=0}^{d_O-1} \sum_{q=0}^{d_P-1} e^{\ii \frac{2 \pi}{d_O} m r + \ii \frac{2 \pi}{d} r l + \ii \frac{2 \pi}{d_P}  l q} \ket{q}_O\ket{r}_P.
\end{equation}
It is easy to see that the exponent simplifies into
\begin{eqnarray}
    \frac{2 \pi}{d_O} m r + \frac{2 \pi}{d} r l + \frac{2 \pi}{d_P}  l q & \equiv & \frac{2 \pi}{d}(m \, d_P + l)(q \, d_O + r) \nonumber \\
    & \equiv & \frac{2 \pi}{d} j k \quad (\textrm{mod} \ 2 \pi),
\end{eqnarray}
where we defined $\ket{k} = \ket{q \, d_O + r} = \ket{q}_O\ket{r}_P$ in accordance with Eq.~\ref{eq:tensor_index_rev} and we used relation $d = d_O \times d_P$. As a result we obtain the transformation rule
\begin{equation}
    \ket{j} \to \frac{1}{\sqrt{d}}\sum_{j=0}^{d-1} e^{\ii \frac{2 \pi}{d} j k} \ket{k},
\end{equation}
which is equivalent to formula in Eq.~\ref{eq:coef_formula_fft}. This completes the proof.

\section{Optical elements}
\label{sec:optical_elements}

\begin{figure}
    \centering
    \includegraphics[scale=.345]{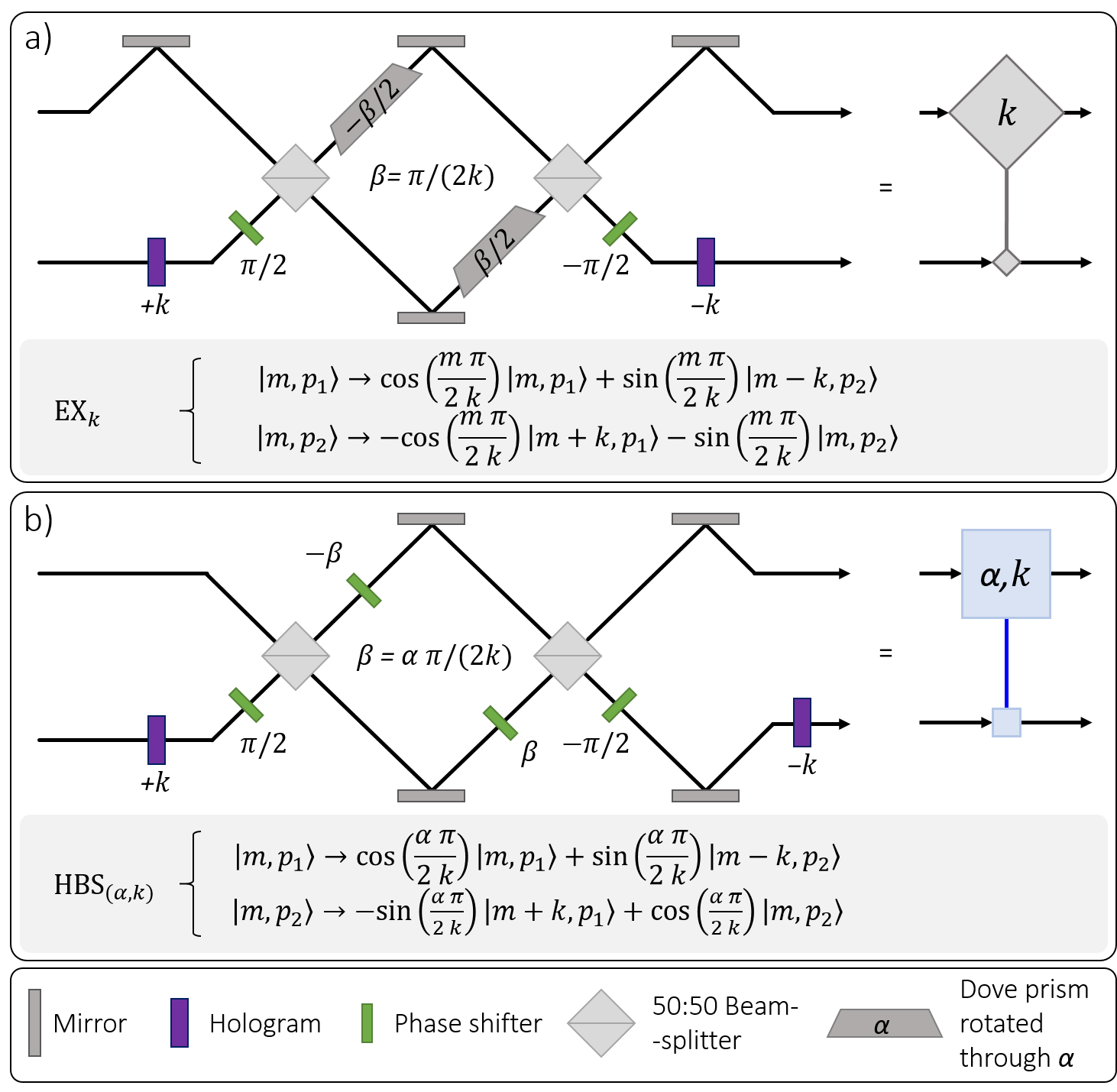}
    \caption{OAM exchangers and holo-beam-splitters. a) The OAM exchanger $\textrm{EX}_k$ of order $k$ is an interferometric device made out of two holograms with opposite values and the OAM parity sorter \cite{Leach2002}. Its operation is captured by the transformation formulas depicted at the bottom. One exchanger of order $k$ effectively works as $4 k$ different beam-splitters with varying splitting ratio for different OAM eigenstates. Note the reversed order of sine and cosine functions in the formula for $\ket{m, p_2}$. b) The holo-beam-splitter $\textrm{HBS}_{(\alpha, k)}$ of order $(\alpha, k)$ comprises a beam-splitter with splitting ratio $\alpha \, \pi / (2 k)$ accompanied by two holograms with opposite values $k$ and $-k$. In the diagram, the implementation of the variable splitting ratio beam-splitter is demonstrated with the help of two 50:50 beam-splitters. The operation of the holo-beam-splitter is captured by the transformation formulas depicted at the bottom.}
    \label{fig:opt_elems}
\end{figure}

\begin{figure}
    \centering
    \includegraphics[scale=.45]{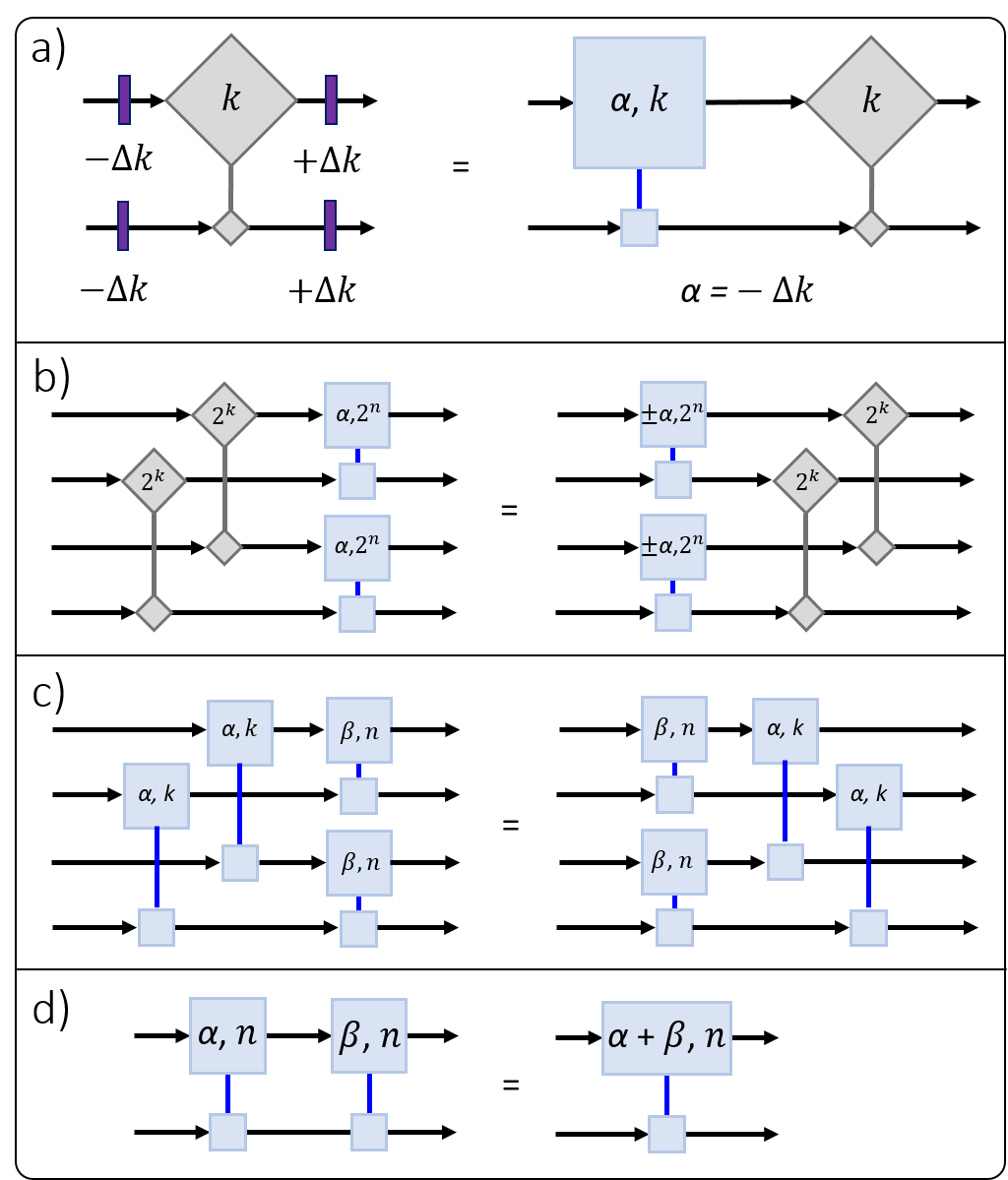}
    \caption{Identities for OAM exchangers and holo-beam-splitters. a) A composition of a holo-beam-splitter of order $(\alpha, k)$ with $\alpha = -\Delta k$ and an OAM exchanger of order $k$ effectively works as a single exchanger of order $k$. This effective exchanger either passes on or reroutes OAM eigenstates of the form $\ket{m \, k + \Delta k}$ that enter the upper port, where $0 < \Delta k < k$ is fixed (and similarly for eigenstates entering the lower port). b) Exchangers of order $2^k$ and holo-beam-splitters of order $(\alpha, 2^n)$ commute in a sense shown in the figure for $n \ge m + 1$. The minus sign is present only for the special case of $n = m + 1$. c) Two pairs of holo-beam-splitters of orders $(\alpha, k)$ and $(\beta, n)$, respectively, commute in a sense shown in the figure. d) Two holo-beam-splitters of orders $(\alpha, n)$ and $(\beta, n)$, respectively, can be combined into a single holo-beam-splitter of order $(\alpha + \beta, n)$.}
    \label{fig:identities}
\end{figure}

The OAM sorter as well as OAM-Path swap consist of two kinds of passive optical elements---OAM exchangers and holo-beam-splitters. Their structure as well as operation are depicted in Fig.~\ref{fig:opt_elems}. Whereas the holo-beam-splitter has a fixed splitting ratio for all OAM eigenstates, the exchanger splits the incoming OAM eigenstates into the two output ports according to a splitting ratio that depends on the OAM eigenstate. As a result, a single exchanger of order $k$ works as $4 k$ different conventional beam-splitters with splitting ratios $0$, $\pi/(2 k)$, $2 \pi/(2 k)$, $\ldots$, $(4 k - 1)\pi/(2 k)$.

The trivial example of an OAM-Path swap is a single exchanger of order $k$. OAM eigenstates $\ket{m \, k}$ entering its input port, where $m \, k$ is a multiple of $k$, are either not affected by the exchanger, or rerouted to the other output port. All remaining OAM eigenstates $\ket{m \, k + \Delta k}$ with $0 < \Delta k < k$ leave the exchanger in a superposition of the two output ports. The same exchanger can be used to reroute such non-multiple OAM eigenstates as well if input eigenstates are first shifted by $-\Delta k$ with a hologram as shown on the left-hand side in Fig.~\ref{fig:identities} a). In such a case though, the superpositions are introduced to the multiple OAM eigenstates instead. As an alternative to such an approach we note that the composition of an exchanger of order $k$ and a holo-beam-splitter of order $(\alpha, k)$ with $\alpha = -\Delta k$ performs the identical rerouting operation, see Fig.~\ref{fig:identities} a). The advantage of the latter approach is that additional exchangers can be inserted between the two elements. These additional exchangers can reroute unwanted terms away from the setup and inject wanted terms in instead. By using this idea iteratively, the OAM-Path swap operator can be constructed.

It is a matter of several simple goniometric transformations to prove the identities demonstrated in Fig.~\ref{fig:identities} b), c) and d). These identities will nonetheless greatly simplify the following discussion of the working principle of the OAM-Path swap operator.

\section{General structure of the OAM-Path swap}
\label{sec:structure_swap}

\begin{figure*}
    \centering
    \includegraphics[scale=.52]{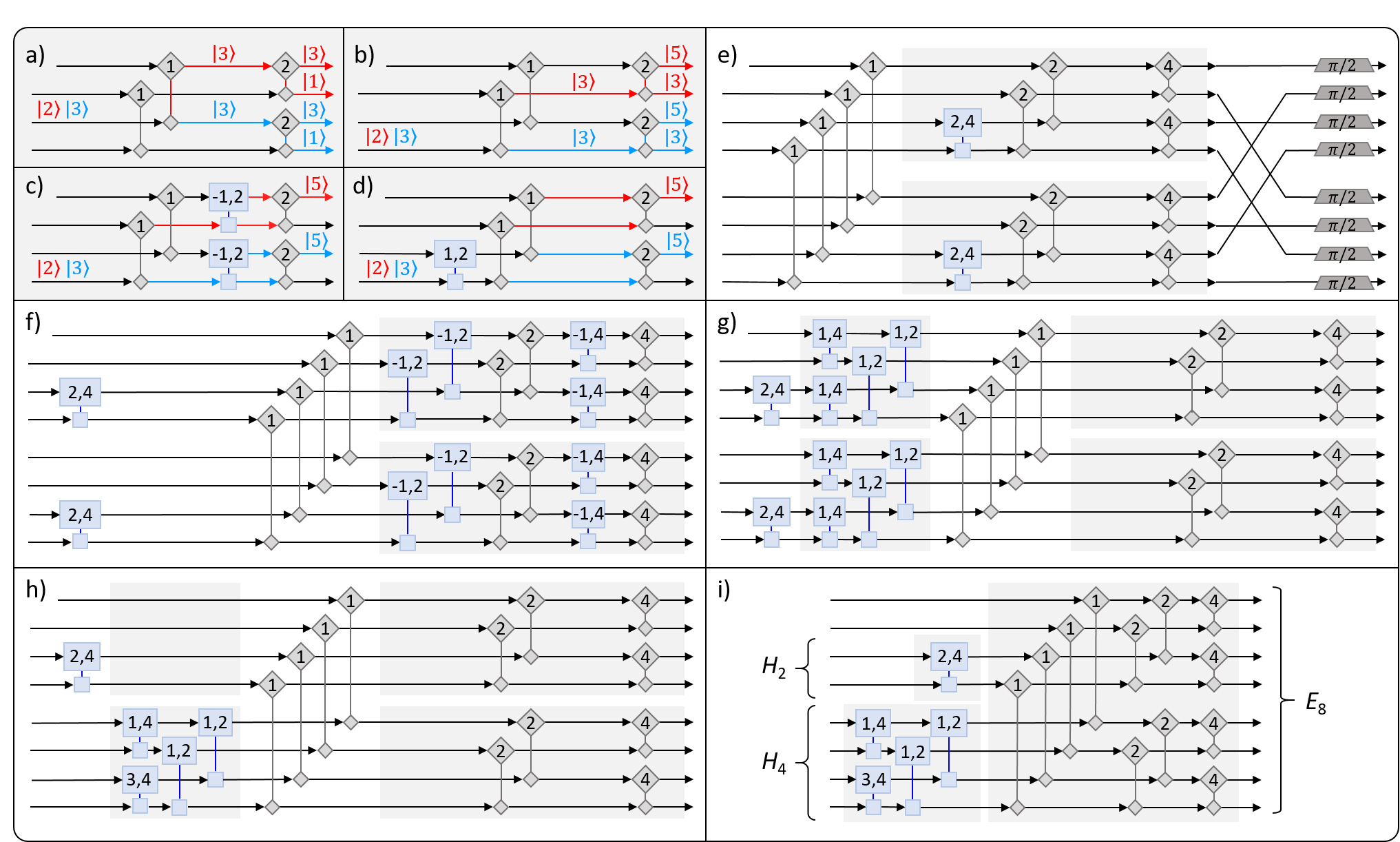}
    \caption{The structure of OAM-Path swap operator. a), b) A specific case of a network of OAM exchangers for $\bar{d} = 4$ and propagation of OAM eigenstates $\ket{2}$ and $\ket{3}$ entering the third and forth paths. The OAM eigenstates leave the network in a superposition. c) When holo-beam-splitters are inserted as suggested in the figure the undesirable splitting of eigenstates is avoided. d) The first column of exchangers can be swapped with the holo-beam-splitters. Futhermore, the upper holo-beam-splitter can be removed such that it does not affect eigenstates injected into the first and second port. As a result, we obtain a setup that sorts correctly all OAM eigenstates injected to any of the input ports. e) The idea of inserting holo-beam-splitters can be generalized to higher dimensions. We demonstrate the general idea for $\bar{d}=2^3$. The setup then consists of two setups for $\bar{d} = 4$ from d) that are connected by four exchangers of order 1. Notice also that orders of exchangers and holo-beam-splitters in the two setups have to be multiplied by two. To comply with the formula in Eq.~\ref{eq:oam_transposition} an additional permutation of paths and a series of properly rotated Dove prisms are necessary in the final part of the setup. As the last part does not alter the evolution of eigenstates through the network we omit it in the following. f) The holo-beam-splitters from the $\bar{d}/2$-dimensional swaps can be moved to the left thanks to identity in Fig.~\ref{fig:identities} b). At this point, all OAM eigenstates that enter any of the first $\bar{d}/2 = 4$ ports are sorted correctly. For the other four input ports the OAM eigenstates undergo more complex evolution and leave the setup in a superposition. In analogy to c), appropriately chosen holo-beam-splitters are inserted into the setup to preclude creation of such superpositions. g) Using identities from Fig.~\ref{fig:identities} b) and c) all holo-beam-splitters can be aggregated in front of the network of exchangers. h) Holo-beam-splitters aggregated this way in the first four paths nevertheless negatively affect OAM eigenstates propagating through these paths. We can remove these holo-beam-splitters to restore the sorting properties of the network for these input paths. In the case of the last four paths, we can use identity from Fig.~\ref{fig:identities} d) to merge all holo-beam-splitters originating in the setup for $\bar{d} = 4$ with the aggregated holo-beam-splitters. For $\bar{d} = 4$ there is only one holo-beam-splitter. We can merge it with its neighbour as made clear in the figure. i) This way we obtained the setup for the OAM-Path swap in $\bar{d} = 8$. All OAM eigenstates that enter any of the eight input ports are sorted correctly. We can identify two conceptually different parts of the swap operator. The network of exchangers, henceforth referred to as an \emph{E block} and a series of networks of increasing size made out of holo-beam-splitters, which we refer to as \emph{H blocks}.
    }
    \label{fig:principle_of_OAM_multiplexer}
\end{figure*}

To illustrate the operating principle of the OAM-Path swap for general dimensions $d_P$ and $d_O$, let us focus first on the simplest case with $d_P = d_O = 4$, which is depicted in Fig.~\ref{fig:principle_of_OAM_multiplexer} a)--d). When the network of OAM exchangers is used to reroute OAM eigenstates entering different input ports, undesirable superpositions of eigenstates are created by higher-order exchangers. All eigenstates entering the first or the second path are rerouted correctly as in the case of the OAM sorter. Nevertheless, for all other paths, the incoming eigenstate leaves the network in a superposition. Utilizing the observation in Fig.~\ref{fig:identities} a) we can insert holo-beam-splitters as demonstrated in Fig.~\ref{fig:principle_of_OAM_multiplexer} c) to fix the undesirable splitting of eigenstates by the last column of exchangers (see also the previous section). Applying identity from Fig.~\ref{fig:identities} b) we can swap the first column of exchangers with the holo-beam-splitters. This way we fixed behavior of exchangers for eigenstates entering the third and forth input ports. Before the addition of holo-beam-splitters, eigenstates injected into the first and second port were rerouted correctly. This property would be destroyed had we left the upper holo-beam-splitter in place. When we remove it and keep only the lower holo-beam-splitter, we arrive at the setup depicted in Fig.~\ref{fig:principle_of_OAM_multiplexer} d), which sorts correctly all OAM eigenstates injected to any of the input ports.

The construction of the OAM-Path swap for general dimensions is recursive and relies heavily on the property illustrated in Fig.~\ref{fig:identities} a). Let us assume for the moment that $d_P = d_O \equiv \bar{d}$. The swap in dimension $\bar{d} = 2^{M}$ is constructed from two swaps in dimension $\bar{d}/2 = 2^{M-1}$ that are connected by a layer of additional exchangers. For a specific example in $\bar{d} = 8$ refer to Fig.~\ref{fig:principle_of_OAM_multiplexer} e), where the $\bar{d}/2$-dimensional swap is presented in Fig.~\ref{fig:principle_of_OAM_multiplexer} d). All OAM eigenstates entering first $\bar{d}/2$ ports are sorted correctly due to properties of $\bar{d}/2$-dimensional swaps. Nevertheless, eigenstates that are injected into the other $\bar{d}/2$ ports leave the network in a superposition of $\bar{d}/2$ output ports. The reason is that the layer of additional exchangers adds one quantum of OAM to such eigenstates and their value is thus no longer a multiple of the order of exchangers in the remaining layers. These exchangers then act not as switches, but rather as genuine beam-splitters.

As follows from Fig.~\ref{fig:identities} a), the switch-like behaviour can be restored if specifically chosen holo-beam-splitters are added to the remaining layers of exchangers, see also Fig.~\ref{fig:principle_of_OAM_multiplexer} f). Making use of identities in Fig.~\ref{fig:identities} b) and c) all such holo-beam-splitters can be aggregated at the beginning of the network as shown in Fig.~\ref{fig:principle_of_OAM_multiplexer} g). At this point, one has to recall that the holo-beam-splitters were introduced to correctly reroute eigenstates entering \emph{last} $\bar{d}/2$ ports. Eigenstates entering \emph{first} $\bar{d}/2$ ports should not be affected by the additional holo-beam-splitters. All such holo-beam-splitters are therefore removed from the first $\bar{d}/2$ paths. This step is illustrated in Fig.~\ref{fig:principle_of_OAM_multiplexer} h). To save resources, one can merge the additional holo-beam-splitters in last $\bar{d}/2$ paths with those holo-beam-splitters that are there due to the $\bar{d}/2$ dimensional swap. At the end, we obtain the OAM-Path swap in dimension $\bar{d}$ that correctly sorts OAM eigenstates injected to any of its $\bar{d}$ input ports as shown in Fig.~\ref{fig:principle_of_OAM_multiplexer} i). The network of exchangers forms the routing part of the swap operator, which we refer to as an \emph{E block}, and the presence of superpositions in output modes is corrected for by a series of networks of increasing size made out of holo-beam-splitters. We refer to these smaller networks as \emph{H blocks}.

In order for the setup to implement the swap transformation in the sense of Eq.~\ref{eq:oam_transposition}, an additional permutation of paths has to be appended to the setup \emph{in each recursion} (cf. Fig.~\ref{fig:principle_of_OAM_multiplexer} e)). This permutation reroutes eigenstates from paths $(p_0, p_1, p_2, \ldots, p_{\bar{d}-1})$ to paths $(p_0, p_2, p_4, \ldots, p_{\bar{d}-2}, p_1, p_3, \ldots, p_{\bar{d}-1})$. As the last stage, a Dove prism rotated through the same angle has to be inserted in each path, compare again with Fig.~\ref{fig:principle_of_OAM_multiplexer} e). These Dove prisms correct for alternating phases of eigenstates leaving the network. Let us note that $H$ blocks can be simplified even more as identity similar to that in Fig.~\ref{fig:identities} b) exists for beam-splitters and holograms. This way, all holograms present in holo-beam-splitters can be put to the sides of the block, whose middle part is then formed merely by conventional beam-splitters.

In cases when $d_O \neq d_P$ one constructs the network for dimension $\bar{d} = \max{(d_O, d_P)}$ and then uses only first $d_P$ input ports and first $d_O$ output ports of the network. Obviously, some elements then do not enter the evolution of injected OAM eigenstates and can be removed from the network with no effect on the swap functionality.

\section{Scaling of the number of elements for high dimensions}
\label{sec:scaling_high_dimensions}

\begin{table*}[!htb]
    \centering
    \begin{tabular}{|p{5ex}|p{5ex} p{5ex} p{5ex}| p{7ex} p{7ex} p{7ex} p{7ex} p{7ex}|p{7ex} p{7ex} p{7ex} p{7ex} p{7ex}|}
    \hline
    & & & & \multicolumn{5}{|c|}{semi-brute-force approach} & \multicolumn{5}{c|}{our approach} \\
 M & $d$ & $d_O$ & $d_P$ & BS & Dove & holo & phas & total & BS & Dove & holo & phas & total \\ \hline
 1 & 2 & 1 & 2 & 5 & 4 & 2 & 2 & 13 & 5 & 4 & 2 & 2 & 13 \\
 2 & 4 & 1 & 4 & 16 & 12 & 6 & 7 & 41 & 16 & 12 & 6 & 7 & 41 \\
 3 & 8 & 2 & 4 & 40 & 28 & 14 & 19 & 101 & 39 & 35 & 20 & 19 & 113 \\
 4 & 16 & 4 & 4 & 92 & 60 & 30 & 47 & 229 & 89 & 75 & 40 & 43 & 247 \\
 5 & 32 & 4 & 8 & 204 & 124 & 62 & 111 & 501 & 181 & 143 & 84 & 87 & 495 \\
 6 & 64 & 4 & 16 & 444 & 252 & 126 & 255 & 1077 & 373 & 279 & 180 & 179 & 1011 \\
 7 & 128 & 8 & 16 & 956 & 508 & 254 & 575 & 2293 & 757 & 667 & 416 & 383 & 2223 \\
 8 & 256 & 16 & 16 & 2044 & 1020 & 510 & 1279 & 4853 & 1597 & 1355 & 768 & 799 & 4519 \\
 9 & 512 & 8 & 64 & 4348 & 2044 & 1022 & 2815 & 10229 & 3093 & 2587 & 1744 & 1551 & 8975 \\
 10 & 1024 & 32 & 32 & 9212 & 4092 & 2046 & 6143 & 21493 & 6205 & 4923 & 3008 & 3055 & 17191 \\
 11 & 2048 & 32 & 64 & 19452 & 8188 & 4094 & 13311 & 45045 & 12317 & 9659 & 5952 & 6015 & 33943 \\
 12 & 4096 & 32 & 128 & 40956 & 16380 & 8190 & 28671 & 94197 & 24605 & 19131 & 11904 & 11967 & 67607 \\
 13 & 8192 & 32 & 256 & 86012 & 32764 & 16382 & 61439 & 196597 & 49309 & 38075 & 23936 & 23935 & 135255 \\
 \hline
\end{tabular}
    \caption{Exact numbers of optical elements necessary to implement the OAM Fourier transform $F^{(d)}_{\mathrm{OAM}}$ for several lowest dimensions of the form $d = 2^{M}$ with corresponding $d_O$ and $d_P$ as determined by the optimization procedure. The numbers of beam-splitters (BS), Dove prisms (Dove), holograms (holo) and phase-shifters (phas) are presented together with their total sum (total) for each dimension. For comparison, the brute-force approach that consists in using the path-only Fourier transform $F^{(d)}_{Path}$ supplemented with two OAM sorters is also presented. As we use the efficient design of $F^{(d)}_{Path}$ \cite{Torma1996, Barak2007}, this approach is referred to as the semi-brute-force approach. One sees that from $d = 8$ onward our approach needs less beam-splitters than the semi-brute-force approach. When the total number of all the optical elements is considered, our approach is more resource-efficient from $d = 32$ onward.}
    \label{tab:num_of_bs}
\end{table*}

The number of optical elements in the setup reflects the setup's complexity. In this section we estimate how this number scales with the dimension $d$ of the OAM Hilbert space. We focus first on the beam-splitters. In our scheme the incoming state is transformed into a state with $d_O$ OAM eigenstates propagating in $d_P$ different paths, such that $d = d_O \times d_P$. The two numbers $d_O$ and $d_P$ are not fixed and their optimal values can be found, for which the number of beam-splitters in the setup is minimal. It turns out that the optimal scenario tends to prefer the choice with $d_O \approx d_P$. Here we assume from the beginning that $d_O = d_P = \sqrt{d}$ in each recursion and study dimensions of the form $d = 2^{2^k}$, where $k \in \mathbb{N}$.

At first, we have to determine the number of beam-splitters required by individual components of the scheme schown in Fig.~\ref{fig:fft}.
For the OAM sorter and the path-only Fourier transform in dimension $d_P$ one obtains $N^{\mathrm{(sort)}}_{BS}(d_P) = 2(d_P-1)$ \cite{Xgate} and $N^{\mathrm{(pFT)}}_{BS}(d_P) = \frac{d_P}{2}\log_2(d_P)$ \cite{Torma1996, Barak2007}, respectively. The OAM-Path swap comprises one $E$ block and a series of $H$ blocks of increasing size. The $E$ block is built out of $\log_2(d_P)$ groups of $k$-order exchangers $\textrm{EX}_k$, where each group contains $d_P/2$ exchangers, cf. Fig.~\ref{fig:leach_plus_bs}~d). The number of conventional beam-splitters implementing the $E$ block is thus equal to $N^{\mathrm{(E)}}_{BS}(d_P) = d_P \log_2(d_P)$. An $H$ block of size $2^k$ has the same structure as the $E$ block, requiring $\frac{2^k}{2} \log(2^k)$ holo-beam-splitters. (In Fig.~\ref{fig:opt_elems}~b) the asymetric beam-splitter is constructed as an interfereometer, here we nevertheless take it as a single element.) The total number of beam-splitters for $H$ blocks is thus
\begin{equation}
N^{\mathrm{(H)}}_{BS}(d_P) = \sum_{k=1}^{\log_2(d_P)-1} \frac{2^k}{2} \log_2(2^k) = \frac{d_P}{2} \log_2(d_P) - d_P + 1.
\end{equation}
To construct an OAM-Path swap in dimension $d_P$ one therefore needs $N^{\mathrm{(swap)}}_{BS}(d_P) = N^{\mathrm{(E)}}_{BS}(d_P) + N^{\mathrm{(H)}}_{BS}(d_P) = \frac{3}{2} d_P \log_2(d_P) - d_P + 1$ beam-splitters.

The number of conventional beam-splitters used in our implementation of the Fourier transform for dimensions of the form $d_k := 2^{2^k}$ is thus
\begin{eqnarray}
N^{\mathrm{(FT)}}_{BS}(d_k) & = & 2 N^{\mathrm{(sort)}}_{BS}(d_{k-1}) + N^{\mathrm{(pFT)}}_{BS}(d_{k-1}) + \nonumber \\
& & + \, N^{\mathrm{(swap)}}_{BS}(d_{k-1}) + d_{k-1} \ N^{\mathrm{(FT)}}_{BS}(d_{k-1}) \nonumber \\   
& = & 2 d_{k-1} \log_2(d_{k-1}) + 3 d_{k-1} - 3 + \nonumber \\
& & + \, d_{k-1} \ N^{\mathrm{(FT)}}_{BS}(d_{k-1}), \label{eq:recursion}
\end{eqnarray}
where $d_O = d_P = \sqrt{d} = d_{k-1}$. Let us define $c_k := 2 \, d_{k} \log_2(d_{k}) + 3 d_{k} - 3$ such that the last formula turns into
\begin{eqnarray}
N^{\mathrm{(FT)}}_{BS}(d_k) & = & c_{k-1} + d_{k-1} \ N^{\mathrm{(FT)}}_{BS}(d_{k-1}).
\end{eqnarray}
When we apply this relation recursively for $k, k-1, \ldots, 1$, the final formula attains the form
\begin{eqnarray}
N^{\mathrm{(FT)}}_{BS}(d_k) & = & \sum_{j=2}^{k} \left( \prod_{l=j}^{k-1} d_l \right) c_{j-1} + \left( \prod_{l=1}^{k-1} d_l \right) N^{\mathrm{(FT)}}_{BS}(d_1) \nonumber \\
& = & \sum_{j=2}^{k} 2^{2^k - 2^j} c_{j-1} + 2^{2^k - 2} N^{\mathrm{(FT)}}_{BS}(d_1) \nonumber \\
& = & 2 d_k \sum_{j=1}^{k-1} \frac{2^j}{2^{2^j}} + 3 \left( \frac{d_k}{4} - 1 \right) + \frac{d_k}{4} N^{\mathrm{(FT)}}_{BS}(d_1). \nonumber
\end{eqnarray}

We can bound the sum in the previous expression as
\begin{equation}
    \sum_{j=1}^{k-1} \frac{2^j}{2^{2^j}} \le \sum_{j=1}^{k-1} \frac{2^j}{2^{2 j}} \le \sum_{j=1}^{\infty} \frac{1}{2^j} = 1,
\end{equation}
and also directly calculate $N^{\mathrm{(FT)}}_{BS}(d_1) = 16$ so that in the end we arrive at
\begin{equation}
N^{\mathrm{(FT)}}_{BS}(d_k) \le \frac{27}{4} d_k = 6.75 \, d_k
\end{equation}
The number of beam-splitters required in our scheme of the Fourier transform therefore scales linearly with the dimension
\begin{equation}
N^{\mathrm{(FT)}}_{BS}(d) \sim O(d),
\end{equation}
when we consider dimensions of the form $d = 2^{2^k}$. The exact total numbers of beam-splitters required by our scheme are shown in Tab.~\ref{tab:num_of_bs} for several lowest dimensions of the general form $d = 2^{M}$. Optimal choices of $d_O$ and $d_P$ for these dimensions were found by an optimization algorithm that searched for a setup with the minimum number of beam-splitters when a given dimension $d$ was fixed. Even in these cases with a general power-of-two dimension the number of beam-splitters still preserves the linear scaling. When fitting the data in Tab.~\ref{tab:num_of_bs} one observes that this number relates to the dimension as
\begin{equation}
    N^{\mathrm{(FT)}}_{BS}(d) \approx 6.037 \, d.
\end{equation}

In complete analogy to Eq.~\eqref{eq:recursion} one can also find a recursive formula for the number of other optical elements present in the setup. Their explicit forms read
\begin{eqnarray}
N^{\mathrm{(FT)}}_{Dove}(d_k) & = & d_{k-1} \log_2(d_{k-1}) + 6 d_{k-1} - 5 + \nonumber \\
& & + \, d_{k-1} \ N^{\mathrm{(FT)}}_{Dove}(d_{k-1}), \\
N^{\mathrm{(FT)}}_{holo}(d_k) & = & 2 d_{k-1} \log_2(d_{k-1}) + \nonumber \\
& & + \, d_{k-1} \ N^{\mathrm{(FT)}}_{holo}(d_{k-1}), \\
N^{\mathrm{(FT)}}_{phas}(d_k) & = & \frac{3}{2} d_{k-1} \log_2(d_{k-1}) + d_{k-1} - 1 + \nonumber \\
& & + \, d_{k-1} \ N^{\mathrm{(FT)}}_{phas}(d_{k-1}),
\end{eqnarray}
where $N^{\mathrm{(FT)}}_{Dove}$, $N^{\mathrm{(FT)}}_{holo}$ and $N^{\mathrm{(FT)}}_{phas}$ stand for the number of Dove prisms, holograms and phase-shifters, respectively. After performing the corresponding calculations it turns out that the total numbers of Dove prisms, holograms and phase-shifters scale linearly with the dimension
\begin{eqnarray}
N^{\mathrm{(FT)}}_{Dove}(d_k) & \leq & \frac{67}{12} d_{k} < 5.6 \, d_{k}, \\
N^{\mathrm{(FT)}}_{holo}(d_k) & \leq & \frac{7}{2} d_{k} = 3.5 \, d_{k}, \\
N^{\mathrm{(FT)}}_{phas}(d_k) & \leq &  \frac{7}{2} d_{k} = 3.5 \, d_{k}. \\
\end{eqnarray}
The exact numbers of these elements are listed in Tab.~\ref{tab:num_of_bs} for several lowest dimensions of the form $d = 2^M$. By fitting the data from this table, one obtains the following behaviour
\begin{eqnarray}
N^{\mathrm{(FT)}}_{Dove}(d) & \approx & 4.860 \, d, \\
N^{\mathrm{(FT)}}_{holo}(d) & \approx & 3.037 \, d, \\
N^{\mathrm{(FT)}}_{phas}(d) & \approx & 2.976 \, d. \\
\end{eqnarray}

Let us mention in this context that the Fourier transform in the orbital angular momentum can be also implemented in a different way. In the brute-force approach the OAM Fourier transform is implemented with the help of a $d$-dimensional path-only Fourier transform and two $d$-dimensional OAM sorters, as is explained at the end of section \ref{sec:implementation} in the main text. Since here we deal only with dimensions that are powers of two, we can use the efficient design of Ref.~\cite{Torma1996} to implement the path-only Fourier transform. For comparison, the number of optical elements required in this semi-brute-force approach is also shown in Tab.~\ref{tab:num_of_bs}. This approach needs more beam-splitters than our scheme, especially in high dimensions. When the total number of all the elements in the setup are of concern, both our scheme and the semi-brute-force scheme are initially comparable, but already from $d = 32$ onward our scheme is more resource-efficient.

\bibliography{ref}

\end{document}